\documentclass[twocolumn,showpacs,preprintnumbers,showkeys,amsmath,amssymb]{revtex4}
\usepackage{graphicx}
\usepackage{amssymb}
\usepackage{amsmath}
\usepackage{bm}
\usepackage[flushleft]{caption2}

\begin{document}
\setcounter{page}{1}
\title{Coordinate transformations in quaternion spaces}
\author{Zihua Weng}
\email{xmuwzh@xmu.edu.cn.}
\affiliation{School of Physics and
Mechanical \& Electrical Engineering,
\\Xiamen University, Xiamen 361005, China}

\begin{abstract}
The quaternion spaces can be used to describe the property of
electromagnetic field and gravitational field. In the quaternion
space, some coordinate transformations can be deduced from the
feature of quaternions, including Lorentz transformation and
Galilean transformation etc., when the coordinate system is
transformed into others. And some coordinate transformations with
variable speed of light can be obtained in the electromagnetic field
and gravitational field.
\end{abstract}



\maketitle

\section{INTRODUCTION}

The quaternion was invented by W. R. Hamilton \cite{morita}. He
spent a lot of time on the theoretical analysis of the quaternions,
and tried to apply quaternions to describe different physical
phenomena. Later, J. C. Maxwell in the electromagnetic theory
\cite{grusky} applied the quaternion to describe the properties of
the electromagnetic field \cite{schwartz}.

In the late 20th century, the quaternions have had a revival due to
their utility in describing spatial rotations primarily. The
quaternion representation of rotations are more compact and faster
to compute than the representations by matrices. And they find uses
in both theoretical and applied physics, in particular for
calculations involving three-dimensional rotations, such as in 3D
computer graphics, control theory, signal processing, and orbital
mechanics etc., although they have been superseded in many
applications by vectors and matrices. Moreover, there are other
quaternion theories, such as quaternion quantum mechanics
\cite{adler}, quaternion optics, quaternion relativity theory
\cite{rastall}, etc.

With the property of quaternions, we obtain Galilean transformation
and Lorentz transformation \cite{teli}, and some other
transformations of coordinate system, including the transformations
with the variable speed of light. In the quaternion spaces, the
speed of light will be varied with the electromagnetic field
potential as well as gravitational field potential.

\section{Transformations in the quaternion space}

The electromagnetic theory can be described with the quaternions. In
the treatise on electromagnetic theory, the algebra of quaternion
was first used by J. C. Maxwell to describe the various properties
of the electromagnetic field. At present, the gravitational field
can be described by the algebra of quaternions as well.

\subsection{Coordinate transformation}
In the quaternion space, we have the radius vector $\mathbb{R} =
(r_0 , r_1 , r_2 , r_3)$ , and the basis vector $\mathbb{E}$ = ($1$,
$\emph{\textbf{i}}_1$, $\emph{\textbf{i}}_2$,
$\emph{\textbf{i}}_3$).
\begin{equation}
\mathbb{R} = r_0 + \emph{\textbf{i}}_1 r_1 + \emph{\textbf{i}}_2 r_2
+ \emph{\textbf{i}}_3 r_3
\end{equation}
where, $r_0 = v_0 t$; $t$ denotes the time; $v_0$ is the speed of
gravitational intermediate boson, which is the first part of the
photon.

The physical quantity $\mathbb{D} (d_0 , d_1 , d_2 , d_3)$ in
quaternion space is defined as
\begin{equation}
\mathbb{D} = d_0 + \emph{\textbf{i}}_1 d_1 + \emph{\textbf{i}}_2 d_2
+ \emph{\textbf{i}}_3 d_3
\end{equation}

When we transform one form of the coordinate system into other one,
the physical quantity $\mathbb{D}$ is transformed into $\mathbb{D}'
(d'_0 , d'_1 , d'_2 , d'_3)$.
\begin{equation}
\mathbb{D}' = \mathbb{K}^* \circ \mathbb{D} \circ \mathbb{K}
\end{equation}
where, $\mathbb{K}$ is the quaternion, and $\mathbb{K}^* \circ
\mathbb{K} = 1$ ; $*$ denotes the conjugate of quaternion.

In case of the coordinate system is transforming, the quaternions in
the above satisfy the relation as follows.
\begin{equation}
\mathbb{D}^* \circ \mathbb{D} = (\mathbb{D'})^* \circ \mathbb{D'}
\end{equation}

When the scalar part of quaternion physical quantity $\mathbb{D}$
does not take part in the coordinate transformations, the scalar
part $d_0$ remains the same.
\begin{equation}
d_0 = d'_0
\end{equation}

From Eqs.(4) and (5), we can obtain many kinds of coordinate
transformations in the quaternion space.

\begin{table}[b]
\caption{\label{tab:table1}The quaternion multiplication table.}
\begin{ruledtabular}
\begin{tabular}{ccccc}
$ $ & $1$ & $\emph{\textbf{i}}_1$  & $\emph{\textbf{i}}_2$ &
$\emph{\textbf{i}}_3$  \\
\hline $1$ & $1$ & $\emph{\textbf{i}}_1$  & $\emph{\textbf{i}}_2$ &
$\emph{\textbf{i}}_3$  \\
$\emph{\textbf{i}}_1$ & $\emph{\textbf{i}}_1$ & $-1$ &
$\emph{\textbf{i}}_3$  & $-\emph{\textbf{i}}_2$ \\
$\emph{\textbf{i}}_2$ & $\emph{\textbf{i}}_2$ &
$-\emph{\textbf{i}}_3$ & $-1$ & $\emph{\textbf{i}}_1$ \\
$\emph{\textbf{i}}_3$ & $\emph{\textbf{i}}_3$ &
$\emph{\textbf{i}}_2$ & $-\emph{\textbf{i}}_1$ & $-1$
\end{tabular}
\end{ruledtabular}
\end{table}

\subsection{Galilean transformation}
In the quaternion space, the velocity $\mathbb{V} (v_0 , v_1 , v_2 ,
v_3 )$ is
\begin{equation}
\mathbb{V} = v_0 + \emph{\textbf{i}}_1 v_1 + \emph{\textbf{i}}_2 v_2
+ \emph{\textbf{i}}_3 v_3
\end{equation}

When the coordinate system is transformed into other one, we have a
radius vector $\mathbb{R}' (r'_0 , r'_1 , r'_2 , r'_3 )$ and
velocity $\mathbb{V}' (v'_0 , v'_1 , v'_2 , v'_3 )$ respectively
from Eq.(3).

From Eqs.(1), (3), and (6), we have
\begin{eqnarray}
& r_0 = r'_0\\
& v_0 = v'_0
\end{eqnarray}
and then $(j = 1, 2, 3)$
\begin{eqnarray}
& t_0 = t'_0\\
& \Sigma (r_j)^2 = \Sigma (r'_j)^2
\end{eqnarray}

The above means that if we emphasize especially the important of the
radius vector Eq.(1) and the velocity Eq.(6), we will obtain the
Galilean transformation of the coordinate system from Eqs.(1), (3),
and (6).

\subsection{Lorentz transformation}
The physical quantity $\mathbb{D} (d_0 , d_1 , d_2 , d_3)$ in
quaternion space is defined as
\begin{eqnarray}
\mathbb{D} && = \mathbb{R} \circ \mathbb{R}
\nonumber \\
&& = d_0 + \emph{\textbf{i}}_1 d_1 + \emph{\textbf{i}}_2 d_2 +
\emph{\textbf{i}}_3 d_3
\end{eqnarray}

In the above equation, the scalar part remains the same during the
quaternion coordinate system is transforming. From Eq.(5) and the
above, we have
\begin{eqnarray}
(r_0)^2 - \Sigma(r_j)^2  = (r'_0)^2 - \Sigma (r'_j)^2
\end{eqnarray}

The above means the spacetime interval $d_0$ remains unchanged, when
the coordinate system rotates. From Eqs.(8) and (12), we obtain the
Lorentz transformation of the coordinate system.
\begin{eqnarray}
&(r_0)^2 - \Sigma(r_j)^2  = (r'_0)^2 - \Sigma (r'_j)^2 \nonumber
\\
& v_0 = v'_0  \nonumber
\end{eqnarray}

The above means that the Galilean transformation and Lorentz
transformation of the coordinates depend on the choosing from
different combinations of the basic physical quantities. When $r_0^2
\gg \Sigma(r_j)^2$ and $(r'_0)^2 \gg \Sigma (r'_j)^2$ , we have
$r_0^2 \approx (r'_0)^2$ . And then Eq.(12) is reduced to Eq.(7).

\subsection{Variable speed of light}
The physical quantity $\mathbb{Q} (q_0 , q_1 , q_2 , q_3)$ in
quaternion space is defined as
\begin{eqnarray}
\mathbb{Q} && = \mathbb{V} \circ \mathbb{V}
\nonumber \\
&& = q_0 + \emph{\textbf{i}}_1 q_1 + \emph{\textbf{i}}_2 q_2 +
\emph{\textbf{i}}_3 q_3
\end{eqnarray}

When the coordinate system is transformed into other one, we have
one physical quantity $\mathbb{Q'} (q'_0 , q'_1 , q'_2 , q'_3)$ from
Eq.(3). In the above equation, the scalar part remains the same
during the quaternion coordinate system is transforming. From Eq.(5)
and the above, we have
\begin{equation}
(v_0)^2 - \Sigma (v_j)^2 = (v'_0)^2 - \Sigma (v'_j)^2
\end{equation}

The above equation represents the physical quantity $q_0$ remains
unchanged when the coordinate system rotates.

From Eqs.(7) and (14), we obtain the transformation A with variable
speed of light.
\begin{eqnarray}
&r_0 = r'_0 \nonumber
\\
&(v_0)^2 - \Sigma (v_j)^2 = (v'_0)^2 - \Sigma (v'_j)^2 \nonumber
\end{eqnarray}

From  Eqs.(12) and (14), we obtain the transformation B with
variable speed of light.
\begin{eqnarray}
&(r_0)^2 - \Sigma(r_j)^2  = (r'_0)^2 - \Sigma (r'_j)^2 \nonumber
\\
&(v_0)^2 - \Sigma (v_j)^2 = (v'_0)^2 - \Sigma (v'_j)^2 \nonumber
\end{eqnarray}

When $v_0^2 \gg \Sigma(v_j)^2$ and $(v'_0)^2 \gg \Sigma (v'_j)^2$ ,
we obtain $v_0^2 \approx (v'_0)^2$ . And then Eq.(14) is reduced to
Eq.(8).

In a similar way, the quantity $\mathbb{D}$ and $\mathbb{Q}$ can be
defined as other kinds of functions of radius vector $\mathbb{R}$ or
velocity $\mathbb{V}$, such as $\mathbb{D} = \mathbb{R} \circ
\mathbb{R} \circ \mathbb{R} \circ \mathbb{R}$ or $\mathbb{Q} =
\mathbb{V} \circ \mathbb{V} \circ \mathbb{V}$, etc. And then we have
some kinds of complicated coordinate transformations in the
quaternion spaces.

\begin{table}[t]
\caption{\label{tab:table1}Some coordinate transformations in the
quaternion space.}
\begin{ruledtabular}
\begin{tabular}{cc}

$ transformations $ & $ radius~ vector~\&~velocity $ \\
\hline
$Galilean $         & $r_0 = r'_0~,~ \Sigma (r_j)^2 = \Sigma (r'_j)^2 $ \\
$ $                 & $v_0 = v'_0$ \\
\hline
$Lorentz  $         & $(r_0)^2 - \Sigma (r_j)^2 = (r'_0)^2 - \Sigma (r'_j)^2 $ \\
$ $                 & $v_0 = v'_0$ \\
\hline
$transformation~A$  & $r_0 = r'_0~,~ \Sigma (r_j)^2 = \Sigma (r'_j)^2 $ \\
$ $                 & $(v_0)^2 - \Sigma (v_j)^2 = (v'_0)^2 - \Sigma (v'_j)^2$ \\
\hline
$transformation~B$  & $(r_0)^2 - \Sigma (r_j)^2 = (r'_0)^2 - \Sigma (r'_j)^2 $ \\
$ $                 & $(v_0)^2 - \Sigma (v_j)^2 = (v'_0)^2 - \Sigma (v'_j)^2$ \\
\hline
$others$            & $\mathbb{D} = \mathbb{R} \circ \mathbb{R} \circ \mathbb{R} \circ \mathbb{R}~,~etc. $ \\
$ $                 & $\mathbb{Q} = \mathbb{V} \circ \mathbb{V} \circ \mathbb{V}~,~etc. $ \\
\end{tabular}
\end{ruledtabular}
\end{table}

\section{Transformations in the octonion space}

The gravitational field and electromagnetic field both can be
demonstrated by quaternions, but they are quite different from each
other indeed. We add another four-dimensional basis vector to the
ordinary four-dimensional basis vector to include the feature of the
gravitational and electromagnetic fields \cite{weng}.

\subsection{Coordinate transformation}
The basis vector of the quaternion space for the gravitational field
is $\mathbb{E}_g$ = ($1$, $\emph{\textbf{i}}_1$,
$\emph{\textbf{i}}_2$, $\emph{\textbf{i}}_3$), and that for the
electromagnetic field is $\mathbb{E}_e$ = ($\emph{\textbf{I}}_0$,
$\emph{\textbf{I}}_1$, $\emph{\textbf{I}}_2$,
$\emph{\textbf{I}}_3$). The $\mathbb{E}_e$ is independent of the
$\mathbb{E}_g$, with $\mathbb{E}_e$ = ($1$, $\emph{\textbf{i}}_1$,
$\emph{\textbf{i}}_2$, $\emph{\textbf{i}}_3$) $\circ$
$\emph{\textbf{I}}_0$. The basis vectors $\mathbb{E}_g$ and
$\mathbb{E}_e$ can be combined together to become the basis vector
$\mathbb{E}$ of the octonion space.
\begin{eqnarray}
\mathbb{E} && = \mathbb{E}_g + \mathbb{E}_e
\nonumber\\
&& = (1, \emph{\textbf{i}}_1, \emph{\textbf{i}}_2,
\emph{\textbf{i}}_3, \emph{\textbf{I}}_0, \emph{\textbf{I}}_1,
\emph{\textbf{I}}_2, \emph{\textbf{I}}_3)
\end{eqnarray}

The radius vector $\mathbb{R} (r_0 , r_1 , r_2 , r_3 , R_0 , R_1 ,
R_2 , R_3 )$ in the octonion space is
\begin{eqnarray}
\mathbb{R} = && r_0 + \emph{\textbf{i}}_1 r_1 + \emph{\textbf{i}}_2
r_2 + \emph{\textbf{i}}_3 r_3
\nonumber\\
&& + \emph{\textbf{I}}_0 R_0 + \emph{\textbf{I}}_1 R_1 +
\emph{\textbf{I}}_2 R_2 + \emph{\textbf{I}}_3 R_3
\end{eqnarray}
and the velocity $\mathbb{V} (v_0 , v_1 , v_2 , v_3 , V_0 , V_1 ,
V_2 , V_3 )$  is
\begin{eqnarray}
\mathbb{V} = && v_0 + \emph{\textbf{i}}_1 v_1 + \emph{\textbf{i}}_2
v_2 + \emph{\textbf{i}}_3 v_3
\nonumber\\
&& + \emph{\textbf{I}}_0 V_0 + \emph{\textbf{I}}_1 V_1 +
\emph{\textbf{I}}_2 V_2 + \emph{\textbf{I}}_3 V_3
\end{eqnarray}
where, $R_0 = V_0 T$; $T$ is one time-like quantity; $V_0$ is the
speed of electromagnetic intermediate boson, which is the second
part of the photon.

When the coordinate system is transformed into other one, the
octonion physical quantity $\mathbb{D}$ will be transformed into
$\mathbb{D}' (d'_0 , d'_1 , d'_2 , d'_3 , D'_0 , D'_1 , D'_2 , D'_3
)$ .
\begin{equation}
\mathbb{D}' = \mathbb{K}^* \circ \mathbb{D} \circ \mathbb{K}
\end{equation}
where, $\mathbb{K}$ is the octonion, and $\mathbb{K}^* \circ
\mathbb{K} = 1$; $*$ denotes the conjugate of octonion.

In case of the coordinate system is transforming, the octonions in
the above satisfy the relation as follows.
\begin{eqnarray}
\mathbb{D}^* \circ \mathbb{D} = (\mathbb{D'})^* \circ \mathbb{D'}
\end{eqnarray}

When the scalar part of octonion does not take part in the
coordinate transformation, the $d_0$ remains the same.
\begin{equation}
d_0 = d'_0
\end{equation}

From Eqs.(19) and (20), we can obtain many kinds of coordinate
transformations in the octonion space.

\begin{table}[t]
\caption{\label{tab:table1}The octonion multiplication table.}
\begin{ruledtabular}
\begin{tabular}{ccccccccc}
$ $ & $1$ & $\emph{\textbf{i}}_1$  & $\emph{\textbf{i}}_2$ &
$\emph{\textbf{i}}_3$  & $\emph{\textbf{I}}_0$  &
$\emph{\textbf{I}}_1$
& $\emph{\textbf{I}}_2$  & $\emph{\textbf{I}}_3$  \\
\hline $1$ & $1$ & $\emph{\textbf{i}}_1$  & $\emph{\textbf{i}}_2$ &
$\emph{\textbf{i}}_3$  & $\emph{\textbf{I}}_0$  &
$\emph{\textbf{I}}_1$
& $\emph{\textbf{I}}_2$  & $\emph{\textbf{I}}_3$  \\
$\emph{\textbf{i}}_1$ & $\emph{\textbf{i}}_1$ & $-1$ &
$\emph{\textbf{i}}_3$  & $-\emph{\textbf{i}}_2$ &
$\emph{\textbf{I}}_1$
& $-\emph{\textbf{I}}_0$ & $-\emph{\textbf{I}}_3$ & $\emph{\textbf{I}}_2$  \\
$\emph{\textbf{i}}_2$ & $\emph{\textbf{i}}_2$ &
$-\emph{\textbf{i}}_3$ & $-1$ & $\emph{\textbf{i}}_1$  &
$\emph{\textbf{I}}_2$  & $\emph{\textbf{I}}_3$
& $-\emph{\textbf{I}}_0$ & $-\emph{\textbf{I}}_1$ \\
$\emph{\textbf{i}}_3$ & $\emph{\textbf{i}}_3$ &
$\emph{\textbf{i}}_2$ & $-\emph{\textbf{i}}_1$ & $-1$ &
$\emph{\textbf{I}}_3$  & $-\emph{\textbf{I}}_2$
& $\emph{\textbf{I}}_1$  & $-\emph{\textbf{I}}_0$ \\
\hline $\emph{\textbf{I}}_0$ & $\emph{\textbf{I}}_0$ &
$-\emph{\textbf{I}}_1$ & $-\emph{\textbf{I}}_2$ &
$-\emph{\textbf{I}}_3$ & $-1$ & $\emph{\textbf{i}}_1$
& $\emph{\textbf{i}}_2$  & $\emph{\textbf{i}}_3$  \\
$\emph{\textbf{I}}_1$ & $\emph{\textbf{I}}_1$ &
$\emph{\textbf{I}}_0$ & $-\emph{\textbf{I}}_3$ &
$\emph{\textbf{I}}_2$  & $-\emph{\textbf{i}}_1$
& $-1$ & $-\emph{\textbf{i}}_3$ & $\emph{\textbf{i}}_2$  \\
$\emph{\textbf{I}}_2$ & $\emph{\textbf{I}}_2$ &
$\emph{\textbf{I}}_3$ & $\emph{\textbf{I}}_0$  &
$-\emph{\textbf{I}}_1$ & $-\emph{\textbf{i}}_2$
& $\emph{\textbf{i}}_3$  & $-1$ & $-\emph{\textbf{i}}_1$ \\
$\emph{\textbf{I}}_3$ & $\emph{\textbf{I}}_3$ &
$-\emph{\textbf{I}}_2$ & $\emph{\textbf{I}}_1$  &
$\emph{\textbf{I}}_0$  & $-\emph{\textbf{i}}_3$
& $-\emph{\textbf{i}}_2$ & $\emph{\textbf{i}}_1$  & $-1$ \\
\end{tabular}
\end{ruledtabular}
\end{table}

\subsection{Galilean transformation}
When the coordinate system is rotated, we have one radius vector
$\mathbb{R}' (r'_0 , r'_1 , r'_2 , r'_3 , R'_0 , R'_1 , R'_2 , R'_3
)$ and velocity $\mathbb{V}' (v'_0 , v'_1 , v'_2 , v'_3 , V'_0 ,
V'_1 , V'_2 , V'_3 )$ respectively from Eq.(18).

From Eqs.(16), (17), and (18), we have
\begin{eqnarray}
&& r_0 = r'_0\\
&& v_0 = v'_0
\end{eqnarray}
and then $(i = 0, 1, 2, 3)$
\begin{eqnarray}
& t_0 = t'_0
\\
& \Sigma (r_j)^2 + \Sigma (R_i)^2 = \Sigma (r'_j)^2 + \Sigma
(R'_i)^2
\end{eqnarray}

The above means that if we emphasize especially the important of the
radius vector Eq.(16) and the velocity Eq.(17), we obtain the
Galilean transformation of the coordinate system from Eqs.(16),
(17), and (18).

The above states also that the $r_0$ remains unchanged when the
coordinate system rotates, but the $R_0$ keeps changed as a
vectorial component. When $R_i \approx R'_i \approx 0$, Eq.(24) is
reduced to Eq.(10).

\subsection{Lorentz transformation}
The physical quantity $\mathbb{D} (d_0 , d_1 , d_2 , d_3 , D_0 , D_1
, D_2 , D_3 )$ in the octonion space is defined as
\begin{eqnarray}
\mathbb{D} = && \mathbb{R} \circ \mathbb{R}
\nonumber\\
= && d_0 + \emph{\textbf{i}}_1 d_1 + \emph{\textbf{i}}_2 d_2 +
\emph{\textbf{i}}_3 d_3
\nonumber\\
&& + \emph{\textbf{I}}_0 D_0 + \emph{\textbf{I}}_1 D_1 +
\emph{\textbf{I}}_2 D_2 + \emph{\textbf{I}}_3 D_3
\end{eqnarray}

By Eqs.(20) and (25), we have
\begin{eqnarray}
(r_0)^2 - \Sigma (r_j)^2 - \Sigma (R_i)^2 = (r'_0)^2 - \Sigma
(r'_j)^2 - \Sigma (R'_i)^2
\end{eqnarray}

The above represents that the spacetime interval $d_0$ keeps
unchanged when the coordinate system rotates in the octonion space.
When the octonion space is reduced to the quaternion space, the
above equation should be reduced to Eq.(12) in the quaternion space.
By Eqs.(22) and (26), we have Lorentz transformation.
\begin{eqnarray}
& (r_0)^2 - \Sigma (r_j)^2 - \Sigma (R_i)^2 = (r'_0)^2 - \Sigma
(r'_j)^2 - \Sigma (R'_i)^2  \nonumber
\\
& v_0 = v'_0  \nonumber
\end{eqnarray}

In the octonion space, when $R_i \approx R'_i \approx 0$, Eq.(26) is
reduced to Eq.(12) in the quaternion space.

\subsection{Variable speed of light}
The physical quantity $\mathbb{Q}(q_0 , q_1 , q_2 , q_3 , Q_0 , Q_1
, Q_2 , Q_3)$ in octonion space is defined as
\begin{eqnarray}
\mathbb{Q} = && \mathbb{V} \circ \mathbb{V}
\nonumber \\
= && q_0 + \emph{\textbf{i}}_1 q_1 + \emph{\textbf{i}}_2 q_2 +
\emph{\textbf{i}}_3 q_3
\nonumber \\
&& + \emph{\textbf{I}}_0 Q_0 + \emph{\textbf{I}}_1 Q_1 +
\emph{\textbf{I}}_2 Q_2 + \emph{\textbf{I}}_3 Q_3
\end{eqnarray}

When the coordinate system is rotated, we have one physical quantity
$\mathbb{Q'}(q'_0 , q'_1 , q'_2 , q'_3 , Q'_0 , Q'_1 , Q'_2 , Q'_3)$
from Eq.(18). In the above, the scalar part remains the same during
the octonion coordinate system is transforming. From Eq.(18) and the
above, we have
\begin{equation}
(v_0)^2 - \Sigma (v_j)^2 - \Sigma (V_i)^2 = (v'_0)^2 - \Sigma
(v'_j)^2 - \Sigma (V'_i)^2
\end{equation}

The above equation represents the physical quantity $q_0$ remains
unchanged when the coordinate system rotates.

From Eq.(21) and (28), we obtain the transformation A with variable
speed of light.
\begin{eqnarray}
&r_0 = r'_0  \nonumber
\\
&(v_0)^2 - \Sigma (v_j)^2 - \Sigma (V_i)^2 = (v'_0)^2 - \Sigma
(v'_j)^2 - \Sigma (V'_i)^2  \nonumber
\end{eqnarray}

From Eq.(26) and (28), we obtain the transformation B with variable
speed of light.
\begin{eqnarray}
& (r_0)^2 - \Sigma (r_j)^2 - \Sigma (R_i)^2 = (r'_0)^2 - \Sigma
(r'_j)^2 - \Sigma (R'_i)^2  \nonumber
\\
&(v_0)^2 - \Sigma (v_j)^2 - \Sigma (V_i)^2 = (v'_0)^2 - \Sigma
(v'_j)^2 - \Sigma (V'_i)^2  \nonumber
\end{eqnarray}

In a similar way, we have other kinds of the coordinate
transformations in the octonion space. In the octonion space, when
$V_i \approx V'_i \approx 0$, Eq.(28) is reduced to Eq.(14) in the
quaternion space.

\section{Transformations in the octonion compounding space}

In the gravitational field and electromagnetic field demonstrated by
quaternions, the vector radius $\mathbb{R}$ will be extended to
$\mathbb{R}+k_{rx}\mathbb{X}$ to cover the different definitions of
energy. And the octonion space with $\mathbb{R}$ is extended to the
octonion compounding space with $\mathbb{R}+k_{rx}\mathbb{X}$,
although their basis vector $\mathbb{E}$ remains the same.

\subsection{Coordinate transformation}
In the octonion compounding space, the basis vector
\begin{eqnarray}
\mathbb{E} = (1, \emph{\textbf{i}}_1, \emph{\textbf{i}}_2,
\emph{\textbf{i}}_3, \emph{\textbf{I}}_0, \emph{\textbf{I}}_1,
\emph{\textbf{I}}_2, \emph{\textbf{I}}_3)
\nonumber
\end{eqnarray}
and the radius vector $\mathbb{R}$ will be extended.
\begin{eqnarray}
\mathbb{R} \rightarrow ~ \mathbb{R} + k_{rx} \mathbb{X}
\end{eqnarray}
where, $\mathbb{X}(x_0 , x_1 , x_2 , x_3 , X_0 , X_1 , X_2 , X_3 )$
is the octonion.

Therefore, the components of the radius vector $\mathbb{R}$ in
Eq.(16) will be extended,
\begin{eqnarray}
r_i~ \rightarrow ~r_i + k_{rx} x_i~,~ R_i~ \rightarrow ~R_i + k_{rx}
X_i~;
\end{eqnarray}
and that of the velocity $\mathbb{V}$ in Eq.(17) will be extended.
\begin{eqnarray}
v_i~ \rightarrow ~v_i + k_{rx} a_i~,~ V_i~ \rightarrow ~V_i + k_{rx}
A_i~.
\end{eqnarray}
where, $\mathbb{A}(a_0 , a_1 , a_2 , a_3 , A_0 , A_1 , A_2 , A_3 )$
is the potential of gravitational field and electromagnetic field;
$x_0 = a_0 t $; $k_{rx}$ is one coefficient, and $k_{rx} = 1 / v_0$
.

When the coordinate system is transformed into other one, the
octonion $\mathbb{D}$ in the compounding space will be transformed
into $\mathbb{D}' (d'_0 , d'_1 , d'_2 , d'_3 , D'_0 , D'_1 , D'_2 ,
D'_3 )$.
\begin{equation}
\mathbb{D}' = \mathbb{K}^* \circ \mathbb{D} \circ \mathbb{K}
\end{equation}
where, $\mathbb{K}$ is the octonion, and $\mathbb{K}^* \circ
\mathbb{K} = 1$; $*$ denotes the conjugate of octonion.

In case of the coordinate system is transforming, the octonions in
the above satisfy the relation as follows.
\begin{eqnarray}
\mathbb{D}^* \circ \mathbb{D} = (\mathbb{D'})^* \circ \mathbb{D'}
\end{eqnarray}

When the scalar part of physical quantity $\mathbb{D}$ does not take
part in the coordinate transformation, the scalar part $d_0$ remains
the same.
\begin{equation}
d_0 = d'_0
\end{equation}

We can obtain some coordinate transformations in the octonion
compounding space from Eqs.(33) and (34).

\subsection{Galilean transformation}
When the coordinate system is rotated, we have one radius vector
$\mathbb{R}' (r'_0 , r'_1 , r'_2 , r'_3 , R'_0 , R'_1 , R'_2 , R'_3
)$ and velocity $\mathbb{V}' (v'_0 , v'_1 , v'_2 , v'_3 , V'_0 ,
V'_1 , V'_2 , V'_3 )$ respectively from Eq.(32).

In the same way, we have correspondingly the new physical quantity
$\mathbb{X}' (x'_0 , x'_1 , x'_2 , x'_3 , X'_0 , X'_1 , X'_2 , X'_3
)$ and the potential $\mathbb{A}' (a'_0 , a'_1 , a'_2 , a'_3 , A'_0
, A'_1 , A'_2 , A'_3 )$.

From Eqs.(30), (31), and (32), we have
\begin{eqnarray}
&& r_0 + k_{rx} x_0 = r'_0 + k_{rx} x'_0
\\
&& v_0 + k_{rx} a_0 = v'_0 + k_{rx} a'_0
\end{eqnarray}
and then
\begin{eqnarray}
& t_0 = t'_0
\\
& \Sigma (r_j + k_{rx} x_j)^2 + \Sigma (R_i + k_{rx} X_i)^2
\nonumber\\
= & \Sigma (r'_j + k_{rx} x'_j)^2 + \Sigma (R'_i + k_{rx} X'_i)^2
\end{eqnarray}

The above means that if we emphasize especially the important of the
radius vector Eq.(30) and the velocity Eq.(31), we obtain the
Galilean transformation of the coordinate system from Eqs.(30),
(31), and (32).

The above states also that the $(r_0+k_{rx}a_0)$ remains unchanged
when the coordinate system rotates, but the speed of light, $v_0$,
will be changed with the potential, $a_0$, of the gravitational
field from Eq.(36).

\subsection{Lorentz transformation}
The physical quantity $\mathbb{D} (d_0 , d_1 , d_2 , d_3 , D_0 , D_1
, D_2 , D_3 )$ in the octonion space is defined as
\begin{eqnarray}
\mathbb{D} = && (\mathbb{R} + k_{rx} \mathbb{X}) \circ (\mathbb{R} +
k_{rx} \mathbb{X})
\nonumber\\
= && d_0 + \emph{\textbf{i}}_1 d_1 + \emph{\textbf{i}}_2 d_2 +
\emph{\textbf{i}}_3 d_3
\nonumber\\
&& + \emph{\textbf{I}}_0 D_0 + \emph{\textbf{I}}_1 D_1 +
\emph{\textbf{I}}_2 D_2 + \emph{\textbf{I}}_3 D_3
\end{eqnarray}

By Eqs.(30), (31), (32), and Eq.(39), we have
\begin{eqnarray}
&& (r_0 + k_{rx} x_0)^2
\nonumber\\
&& - \Sigma (r_j + k_{rx} x_j)^2 - \Sigma (R_i + k_{rx} X_i)^2
\nonumber\\
= && (r'_0 + k_{rx} x'_0)^2
\nonumber\\
&& - \Sigma (r'_j + k_{rx} x'_j)^2 - \Sigma (R'_i + k_{rx} X'_i)^2
\end{eqnarray}

The above equation states the $d_0$ remains unchanged when the
coordinate system rotates in the octonion compounding space. When
the compounding octonion space is reduced to the compounding
quaternion space, the Eq.(40) will be reduced to that in the latter
space.

The above means also that the speed of light $v_0$ will be changed
with the potential of either the gravitational field or the
electromagnetic field. When the potential $a_i$ and $A_i$ both are
equal approximately to zero, the Eq.(40) will be reduced to Eq.(26).
By Eqs.(36) and (40), we can obtain the Lorentz transformation.

\subsection{Variable speed of light}
The physical quantity $\mathbb{Q}(q_0 , q_1 , q_2 , q_3 , Q_0 , Q_1
, Q_2 , Q_3)$ in octonion space is defined as
\begin{eqnarray}
\mathbb{Q} = && (\mathbb{V}+k_{rx}\mathbb{A}) \circ
(\mathbb{V}+k_{rx}\mathbb{A})
\nonumber \\
= && q_0 + \emph{\textbf{i}}_1 q_1 + \emph{\textbf{i}}_2 q_2 +
\emph{\textbf{i}}_3 q_3
\nonumber \\
&& + \emph{\textbf{I}}_0 Q_0 + \emph{\textbf{I}}_1 Q_1 +
\emph{\textbf{I}}_2 Q_2 + \emph{\textbf{I}}_3 Q_3
\end{eqnarray}

When the coordinate system is rotated, we have a new physical
quantity $\mathbb{Q'}(q'_0 , q'_1 , q'_2 , q'_3 , Q'_0 , Q'_1 , Q'_2
, Q'_3)$ from Eq.(32). In the above equation, the scalar part
remains the same during the octonion coordinate system is
transforming. From Eq.(32) and the above, we have
\begin{eqnarray}
&& (v_0 + k_{rx} a_0)^2
\nonumber\\
&& - \Sigma (v_j + k_{rx} a_j)^2 - \Sigma (V_i + k_{rx} A_i)^2
\nonumber\\
= && (v'_0 + k_{rx} a'_0)^2
\nonumber\\
&& - \Sigma (v'_j + k_{rx} a'_j)^2 - \Sigma (V'_i + k_{rx} A'_i)^2
\end{eqnarray}

The above equation represents the physical quantity $q_0$ remains
unchanged when the coordinate system rotates. From the above and
Eq.(38) or Eq.(40), we obtain the transformations with variable
speed of light.

The above means also that the speed of light $v_0$ will be changed
with either the gravitational potential or the electromagnetic
potential. So we find different kinds of light speed in various
optical waveguide materials. And there exist negative refractive
indexes in optical waveguide materials, when the field potentials
are switched from positive to negative. When the potential $a_i$ and
$A_i$ both are equal approximately to zero, the Eq.(42) will be
reduced to Eq.(28).

\section{CONCLUSIONS}

In the quaternion spaces, Galilean transformation and Lorentz
transformation can be deduced from the feature of quaternions. This
states that Lorentz transformation is only one of several coordinate
transformations in the electromagnetic field and gravitational
field.

In the octonion spaces, there exist some coordinate transformations
with the variable speed of light. In the electromagnetic and
gravitational fields, the gravitational intermediate boson and
electromagnetic intermediate boson can be combined together to
become the photon. So the speed of light will be changed with the
electromagnetic potential and gravitational potential.

It should be noted that the study for the coordinate transformation
has examined only some simple cases, including Galilean
transformation and Lorentz transformation etc. Despite its
preliminary character, this study can clearly indicate that there
are several kinds of coordinate transformations with the variable
speed of light. For the future studies, the investigation will
concentrate on only some suitable predictions about the complicated
coordinate transformations in the electromagnetic field and
gravitational field.

\begin{acknowledgments}
This project was supported partly by the National Natural Science
Foundation of China under grant number 60677039, Science \&
Technology Department of Fujian Province of China under grant number
2005HZ1020 and 2006H0092, and Xiamen Science \& Technology Bureau of
China under grant number 3502Z20055011.
\end{acknowledgments}


\begin{references}


\bibitem{morita}
      Morita K.
      Quaternions, Lorentz Group and the Dirac Theory.
      {\it Progress of Theoretical Physics},
      2007, 117 (3): 501--532.

\bibitem{grusky}
      Grusky S M, Khmelnytskaya K V, Kravchenko V V.
      On a quaternionic Maxwell equation for the time-dependent
           electromagnetic field in a chiral medium.
      {\it Journal of Physics A},
      2004, 37 (16): 4641--4647.

\bibitem{schwartz}
      Schwartz C.
      Relativistic quaternionic wave equation.
      {\it Journal of Mathematical Physics}, 2006, 47
      (12): 122301--122301-13.

\bibitem{adler}
      Adler S L.
      {\it Quaternionic Quantum Mechanics and Quantum Fields},
      (Oxford University Press, Oxford, UK, 2001).

\bibitem{rastall}
      Rastall P.
      Quaternions in Relativity.
      Reviews of Modern Physics,
      1964, 36 (3): 820-832.

\bibitem{teli}
      Teli M T.
      Quaternionic Form of Unified Lorentz Transformations.
      Physics Letters A, 1980, 75 (6): 460--462.

\bibitem{weng}
      Weng Z-H and Weng Y.
      Variation of Gravitational Mass in Electromagnetic Field.
      In: \textit{Proceedings of Progress in Electromagnetics Research Symposium 2009},
      Beijing, China, March 2009, 105--107.


\end{references}
\end{document}